\documentclass[
    aps,prb,twocolumn,
	groupedaddress,superscriptaddress,
	amsfonts,amssymb,amsmath,
	citeautoscript,longbibliography,
	letterpaper, nofootinbib
	]{revtex4-2}

\usepackage{appendix}
\usepackage{graphicx}   
\usepackage{import}                        
\usepackage{epstopdf}
\usepackage{amsmath} 
\usepackage{bm}
\usepackage{amssymb}
\usepackage{quotes}
\usepackage{transparent}
\usepackage{dcolumn}
\usepackage{multirow}
\usepackage{cancel} 
\usepackage{mdframed}
\usepackage{color}
\usepackage{bm}
\usepackage{dsfont}
\usepackage{slashed}
\usepackage{enumitem}
\usepackage{braket}
\usepackage{hyperref}

\renewcommand{\d}{\mathrm{d}}
\newcommand{\im}{\mathrm{i}}
\begin{document}
\rmfamily

\title{Noise immunity in quantum optical systems through non-Hermitian topology}

\author{Jamison Sloan$^\P$}
\email{jsloan@stanford.edu}
\affiliation{E. L. Ginzton Laboratory, Stanford University, Stanford, CA 94305, USA}
\author{Sachin Vaidya$^\P$}
\email{svaidya1@mit.edu}
\affiliation{Department of Physics, Massachusetts Institute of Technology, Cambridge, Massachusetts~02139, USA}
\author{Nicholas Rivera}
\affiliation{School of Applied and Engineering Physics, Cornell University, Ithaca, New York 14853, USA}
\affiliation{Department of Physics, Harvard University, Cambridge MA 02138, USA}
\author{Marin Solja\v{c}i\'{c}}
\affiliation{Research Laboratory of Electronics, Massachusetts Institute of Technology, Cambridge, Massachusetts~02139, USA}
\affiliation{Department of Physics, Massachusetts Institute of Technology, Cambridge, Massachusetts~02139, USA \\
$^\P$ denotes equal contribution}

\renewcommand{\abstractname}{} 
\begin{abstract}
Multimode nonlinear optical systems are highly valued for their ability to withstand large amounts of optical power, transmit data with high bandwidth, perform physical computations, generate quantum correlations, and much more. For many of these applications, both classical and quantum noise place limitations on important performance metrics.
Moreover, it is well known that nonlinear systems tend to develop noise due to their sensitivity to initial conditions, thus motivating the development of general principles which can be used to control and mitigate noise in such systems.  
In this work, we show that non-reciprocity can enable the unidirectional flow of noise from the non-equilibrium steady states of nonlinear driven-dissipative systems. 
We demonstrate that this unidirectional flow results from the non-Hermitian skin effect (NHSE) for quantum noise.
This effect provides a robust mechanism for isolating parts of the system from degradation due to excess injected noise. 
Our findings may lead to novel design principles for multimode amplifiers or oscillators that support high power handling while mitigating the buildup of noise. 
Moreover, our approach provides a new means to connect non-Hermitian topology with nonlinear many-body systems, which is expected to lead to topological descriptions of interacting states of light.
\end{abstract}
\maketitle

\section{Introduction}
Nonlinear dynamical systems are central to scientific and engineering disciplines, and support fascinating universal phenomena such as phase transitions, synchronization, pattern formation, and chaotic behavior \cite{kaplan1997understanding, nayfeh2008applied, thompson1990nonlinear}.
In optics and photonics, nonlinearity forms the backbone of critical applications in frequency conversion, field sensing, ultrashort pulse formation, and quantum state generation \cite{boyd2020nonlinear, shen1984principles, scully1999quantum, shen1967quantum}. 
Over the past few decades, efforts have been directed at understanding and controlling nonlinear optical systems with many degrees of freedom (e.g., frequency/time, space/wavevector, polarization) due to their enhanced capacities to carry power, transmit data, and perform physical computations \cite{wright2022physics, wright2022nonlinear, yanagimoto2024mesoscopic}.
Topics of interest under this broad umbrella include effects in nonlinear waveguide arrays \cite{christodoulides2003discretizing}, controlling nonlinear behaviors in multimode fibers \cite{wright2017spatiotemporal, chen2023mitigating, chen2023suppressing, wu2019thermodynamic, pourbeyram2022direct}, observing quantum effects in soliton microcombs
\cite{guidry2022quantum}, phase transitions in coupled networks of optical parametric oscillators \cite{roy2023non}, and many more. 

A less widely studied---yet highly important---aspect of multimode nonlinear systems is the behavior of \textit{noise}, which limits the performance of many devices and measurement schemes. 
Due to their sensitivity to initial conditions, many multimode nonlinear optical processes (e.g. self- and cross-phase modulation, raman and brillouin scattering) are known to develop excess noise, even when driven by ideal coherent sources. 
Moreover, noise is of central importance in devices which benefit from light with fluctuations below the standard quantum limit (SQL)~\cite{clerk2010introduction}. 
Here, it is well-known that excess noise quickly degrades quantum states (e.g. squeezed state, Fock states, entangled states) through effects which become increasingly difficult to mitigate in large systems. 
This highlights the critical need to develop methods for robustly controlling noise in multimode nonlinear systems.
The identification of new approaches to this challenge---ideally based on robust organizing principles---is vital for advancing scalable and reliable optical technologies such as neural networks~\cite{wright2022deep, wang2023image, onodera2024scaling, presutti2024highly} and sensing schemes~\cite{lau2018fundamental, zhang2019quantum}.

One natural approach to achieving robust physical behaviors is to harness topological effects. 
Over the last decade, such approaches have enabled photonic systems that are resilient to scattering losses and disorder~\cite{ozawa2019topological, smirnova2020nonlinear, lu2014topological}. 
More recently, interest has developed in the interplay between photonic topology and nonlinearity~\cite{ozawa2019topological, smirnova2020nonlinear, schulz2021topological}, leading to explorations in formation and quantized pumping of solitons~\cite{jurgensen2021quantized, jurgensen2022chern, fu2022nonlinear, mukherjee2021observation, de2024solitons}, topological lasing~\cite{bandres2018topological, klembt2018exciton, leefmans2024topological, zhu2022anomalous}, nonlinear control of topological states~\cite{dobrykh2018nonlinear, arkhipova2022observation, jorg2024optical, pontula2024non} and formation of strongly interacting states of light~\cite{clark2020observation, wang2024realization}. 
Emerging research has also begun to explore topological effects in quantum optics. 
For example, topologically protected edge states offer potential as noise-resilient quantum amplifiers and frequency converters~\cite{busnaina2024quantum, mcdonald2020exponentially, mcdonald2022nonequilibrium, peano2016topological, mcdonald2018phase, sulimany2024quantum, wan2023quantum}.
Notably, these topological approaches can relax the configuration sensitivity of existing noise mitigation strategies, rendering the quantum noise properties of these systems globally robust and insensitive to fine details. 

In this work, we present a novel effect in which noise flows unidirectionally in a multimode nonlinear quantum system, achieved through non-reciprocity. 
This non-reciprocity, introduced via asymmetric non-Hermitian couplings between modes, leads remarkably to the complete immunity of certain parts of the system to excess noise, even preserving squeezing in the presence of significant input noise. 
Importantly, this effect is robust to disorder and does not rely on fine-tuning of system parameters, making it broadly applicable across a range of implementations. 
We identify this noise immunity as a topological effect—specifically, a manifestation of the non-Hermitian skin effect (NHSE), which arises from the point-gap topology of the Hamiltonian governing the system's noise. 
Crucially, this behavior of noise flow and immunity is distinct from, and decoupled from, the dynamics of the mean field, which can exhibit a complex interplay between the nonlinearity and non-reciprocity. 
We further uncover novel topological transitions mediated by nonlinearity, alongside states which support ordered quantum correlations inherent to non-Hermitian systems. 
Finally, we discuss how these effects can be realized at optical frequencies in existing platforms including fiber and integrated photonics.

\section{Quantum optics of nonlinear non-reciprocal systems} 
\begin{figure}
    \centering
    \includegraphics[width=\linewidth]{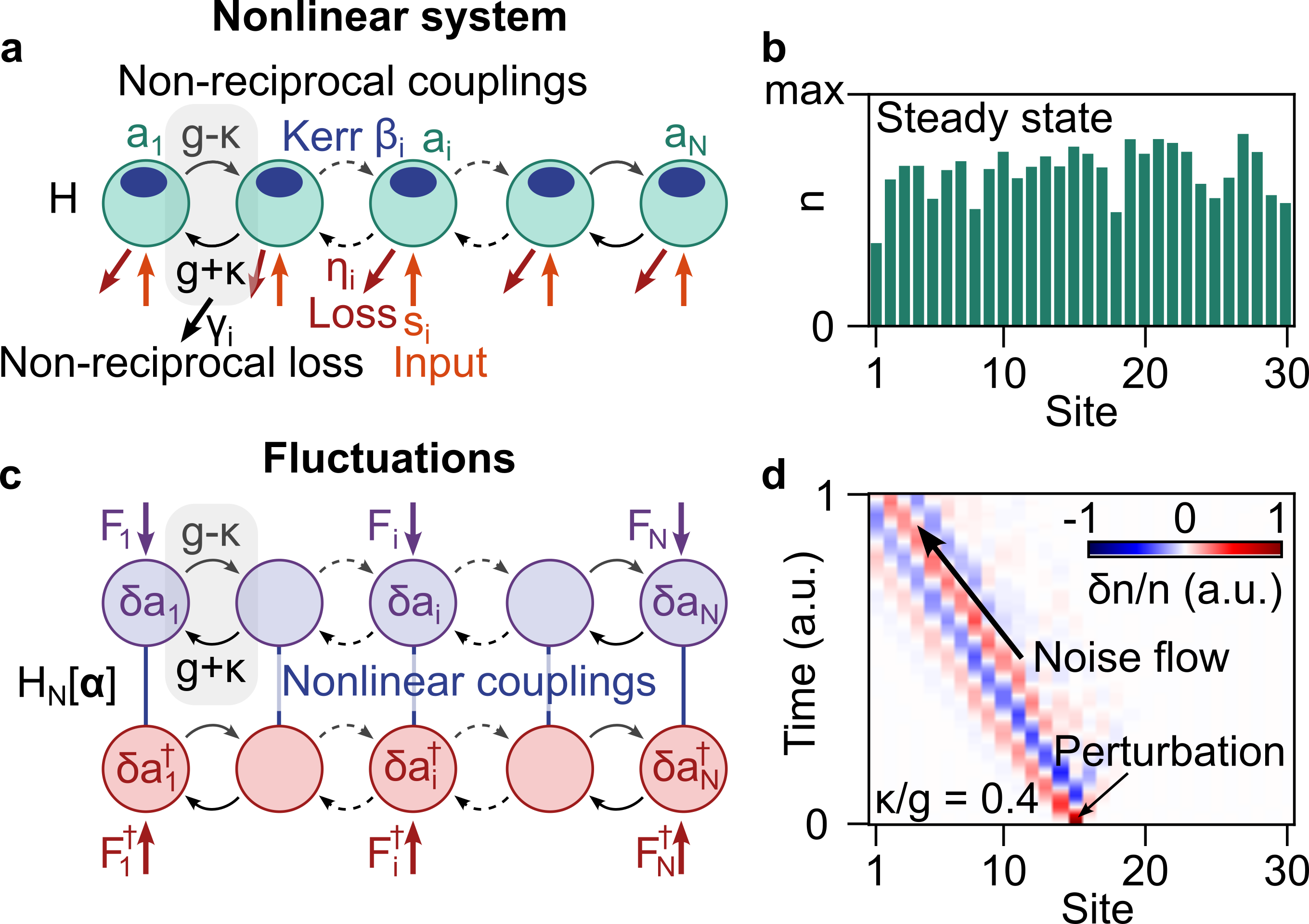}
    \caption{\textbf{Non-reciprocal transport of fluctuations from steady state in a driven-dissipative nonlinear system.} (\textbf{a}) Driven-dissipative Hatano-Nelson model with on-site Kerr nonlinearities which can support non-equilibrium steady states when driven by external channels. (\textbf{b}) Example of such a steady state in a chain of $N=30$ sites driven with detunings $\Delta_i$ which are randomly sampled from the distribution $\Delta_i/\omega_0 \in [0, 0.4)$ for each site $i$. The combination of linear couplings ($g/\omega_0=0.6$), non-reciprocity ($\kappa/g = 0.4$) and nonlinearity ($\beta|s|^2/\omega_0^2=100$) results in a state distribution which does not bear any salient signature of non-reciprocity. The limiting loss rate is $\eta_i/\omega_0 = 10^{-2}$ for all sites, which is supplemented by the diagonal loss $\gamma_i = 2\kappa$ required by non-reciprocity. (\textbf{c}) Linear fluctuations of the nonlinear system from steady state can be described by a non-Hermitian noise Hamiltonian $H_N$ with $2N$ degrees of freedom. The system is driven by Langevin forces $F_i^{(\dagger)}$ which include the noise sources associated with dissipation, non-reciprocity, etc.
    (\textbf{d}) Non-reciprocal transport of fluctuations seen via the classical time dynamics which result from taking the steady state shown in (b), and then perturbing the amplitude of site $i=15$ by a small amount. The quantity shown is $\delta n_i(t)/n_i(0) = (n_i(t) - n_i(0))/n_i(0)$, giving the fractional change in the photon number $n_i$ at each site relative to its steady state value $n_i(0)$.}
    \label{fig:nonreciprocal_noise_concept}
\end{figure}

We first introduce a minimal model which captures the non-reciprocal quantum noise effects discussed in this work. 
Our model consists of a 1D lattice of $N$ bosonic resonances $a_i$ with frequencies $\omega_i$ that each contain Kerr nonlinearity (i.e. an intensity dependent shift to the resonance frequency). 
Each resonance exchanges photons with its nearest neighbors through a non-reciprocal coupling $g\pm\kappa$, where $\kappa$ induces non-reciprocity. In this scheme, each site in the lattice is coupled to its left neighbor more strongly than its right one (for positive $\kappa$) (Fig.~\ref{fig:nonreciprocal_noise_concept}a). 
Such couplings are characteristic of the Hatano-Nelson model \cite{hatano1996localization}, which exhibits the so called ``non-Hermitian skin effect'' in linear systems, which localizes all eigenstates to an edge under open boundary conditions. 
Finally, we assume that each resonance is coupled to external source fields $s_i$ which oscillate at frequency $\omega_p$, driving the system into a non-equilibrium steady-state.

Under these assumptions, the cavity annihilation operators $a_i$ are governed by a system of Heisenberg-Langevin equations of motion (in units where $\hbar=1$):
\begin{subequations}
\begin{align}
    \dot{a}_i &= -\im\sum_j \tilde{H}_{ij}^{\text{(lin)}}a_j - \im[H_K,a_i]  + \sqrt{2\eta_i} s_i(t) + f_i(t) \label{eq:HL_EOM_general} \\
    &= -(\gamma_i  + \eta_i) a_i -\im\Delta_i a_i -\im\beta_i (a_i^\dagger a_i)a_i \nonumber \\
    &- \im(g -\kappa)a_{i-1} -\im(g +\kappa)a_{i+1} + \sqrt{2\eta_i} s_i(t) + f_i(t). \label{eq:HL_EOM_HN}
\end{align}
\end{subequations}
We have defined $\tilde{H}^{\text{(lin)}}$ as an effective non-Hermitian Hamiltonian which accounts for the linear dynamics (including loss), and contains the Hatano-Nelson type couplings.
The Kerr effect is described by the Hamiltonian $H_K = (\beta_i/2)(a_i^\dagger)^2a_i^2$, where $\beta_i$ is the nonlinear frequency shift per photon. 
Additionally, $\Delta_i = \omega_p - \omega_i$ is the detuning of the pump from each resonance, and $\eta_i$ is the decay rate associated with coupling to the input field $s_i$. 
These input fields have bosonic commutation relations $[s_i(t), s_j(t')] = \delta_{ij}\delta(t-t')$, and for coherent input light, have the correlations $\braket{s_i(t)s_j^\dagger(t')} = \delta_{ij}\delta(t-t')$ (while all other possible correlators vanish). 
In some parts of this work, we will also consider the injection of noisy input light, which requires different correlators (see Supplementary Information (S.I.) for details).

Each field $a_i$ is additionally driven by an operator-valued Langevin force $f_i(t)$ which is required for a quantum mechanically consistent model that includes the non-reciprocity $\kappa$.
This is because the non-reciprocal terms here are necessarily non-Hermitian, and therefore must arise from the coupling of the system fields to some external bath degrees of freedom (i.e. loss and/or gain channels). 
The specifics of these external degrees of freedom do not influence the mean-field behavior of the system, but are crucial to describing quantum noise behaviors. 
In this work, we focus on cases where the non-reciprocity arises purely from coupling to dissipative baths. 
Such non-reciprocal couplings can be achieved dissipatively through schemes such as magneto-optical isolation, or electro-optic modulation schemes \cite{wang2021generating, herrmann2022mirror, orsel2024giant}. 
For dissipative non-reciprocity, the Langevin forces associated with the external loss baths have correlations given by: $\braket{f_i(t) f_i^\dagger(t')} = 2\gamma \delta(t-t')$, and $\braket{f_i(t) f_{i\pm 1}^\dagger(t')} = \pm 2\im\kappa\delta(t-t')$. 
Here, $\gamma$ is an additional on-site loss which must accompany the non-reciprocal couplings.
The minimum allowable value of $\gamma$ is set by the requirement that the matrix of loss bath correlations is positive semi-definite, which for this model means that $\gamma \geq 2\kappa$.
Additional information about quantum noise sources for non-Hermitian Hamiltonians can be found in \cite{mcdonald2022nonequilibrium}, and in the S.I. of this work.

In this work, we are concerned primarily with the quantum fluctuations of the system from some non-equilibrium steady-state which arises from external driving. 
In this case, the annihilation operator of each mode is written as $a_i = \alpha_i + \delta a_i$, where $\alpha_i \equiv \braket{a_i}$ is the steady-state mean-field, and $\delta a_i$ is the operator-valued fluctuation. We will notate such fluctuations with the vector $(\boldsymbol{\delta a}, \boldsymbol{\delta a}^\dagger)^T \equiv (\delta a_1, \ldots \delta a_N, \delta a_1^\dagger, \ldots, \delta a_N^\dagger)^T$.
To analyze noise, we linearize the equations of motion (Eq.~\ref{eq:HL_EOM_general}) around the steady-state values $\boldsymbol{\alpha}$. 
Such an approach is accurate when the quantum fluctuations are small relative to the coherent field amplitudes (as is typically the case in the context of nonlinear optics with macroscopic photon numbers where quantum states remain Gaussian). 
Doing so, we obtain:

\begin{equation}
    \frac{\d}{\d t} \begin{pmatrix}
        \boldsymbol{\delta a} \\ \boldsymbol{\delta a}^\dagger
    \end{pmatrix} = 
    -\im \underbrace{\begin{pmatrix}
        U & V \\
        -V^* & -U^*
    \end{pmatrix}}_{H_N[\boldsymbol{\alpha}]} \begin{pmatrix}
        \boldsymbol{\delta a} \\ \boldsymbol{\delta a}^\dagger
    \end{pmatrix} + \begin{pmatrix}
        \mathbf{F} \\ \mathbf{F}^\dagger
    \end{pmatrix}.
    \label{eq:noise_hamiltonian_eom}
\end{equation}
Here, $H_N[\boldsymbol{\alpha}]$ is a non-Hermitian Hamiltonian for the noise that depends on the steady-state fields, $\boldsymbol{\alpha}$. 
For a system with on-site Kerr nonlinearity, the matrices $U$ and $V$ are given as

\begin{subequations}
\begin{align}
    U_{ij} &= 2\beta_i|\alpha_i|^2\delta_{ij} + \tilde{H}_{ij}^{\text{(lin)}} \label{eq:U} \\
    V_{ij} &= \beta_i\alpha_i^2\delta_{ij}. \label{eq:V}
\end{align}
\end{subequations}
Additionally, $\mathbf{F}= (F_1, \ldots, F_N)^T$ is the set of Langevin forces which drive fluctuations. 
Each total force is given as the sum of the forces which arise from the non-reciprocity and the coupling to the driving fields: $F_i \equiv f_i + \sqrt{2\eta_i}\delta s$, where $\delta s_i \equiv s_i - \braket{s_i}$. 
The couplings of Eq.~\ref{eq:noise_hamiltonian_eom} can be interpreted in terms of a lattice of extended dimension which couples the degrees of freedom $(\boldsymbol{\delta a}, \boldsymbol{\delta a}^\dagger)^T$, while driven by quantum noise sources $\textbf{F}$ (Fig.~\ref{fig:nonreciprocal_noise_concept}c). 
When written in terms of real variables, such a model corresponds to so-called ``quadrature lattices''~\cite{lustig2024emerging, mcdonald2018phase, slim2024optomechanical}. 

In the context of our model, this lattice exhibits two key features: (1) non-reciprocal couplings $g \pm \kappa$ which are inherited from those of the physical system $H$ (see Eq.~\ref{eq:U}), and (2) couplings between fluctuation degrees of freedom $\delta a_i$ and $\delta a_i^\dagger$ induced by the local Kerr nonlinearity $\beta_i$ (see Eq.~\ref{eq:V}). 
We also note that while we focus in this work on models with Hatano-Nelson couplings, the forms of Eqs.~\ref{eq:noise_hamiltonian_eom}-3 apply generally to any model with Kerr nonlinearity and non-Hermitian dynamics governed by $\tilde{H}^{(\text{lin})}$. 
We therefore anticipate that many of the concepts presented here should generalize to more complex non-reciprocal coupling models (e.g. beyond nearest neighbor hopping).

We first demonstrate that $H_N$ can support a non-reciprocal flow of fluctuations from equilibrium, even for a highly nonlinear steady-state. 
To show this, we consider the nonlinear driven-dissipative Hatano-Nelson model with open boundaries driven by a randomly selected set of drive fields $s_i$ and detunings $\Delta_i$. 
We choose random parameters to demonstrate that the new effect we predict does not depend on carefully selected parameters. 
The mean-field steady-state is found by numerically solving the transient mean-field dynamics of the driven system~\footnote{Note that for a given set of system parameters and input fields, there may in principle be multiple steady-states (or no stable steady-state). We work in a regime where stable steady-states exist, and take the one which is found by pumping from the ground state.}. 
The resulting steady-state features a near-uniform distribution of photons across all different lattice sites (Fig.~\ref{fig:nonreciprocal_noise_concept}b), which arises from the competition of two effects: (1) non-reciprocity, which tends to push population toward the left boundary (for $\kappa > 0$), and (2) nonlinearity, which causes more populated sites to detune away from the pump, thus resisting high intensities. 
Hence, while in the linear limit $(\beta = 0)$, the dynamics of excitations in this system is dominated by chiral transport, as is well-known in the Hatano-Nelson model, it is clear that nonlinearities can counteract such behavior. 
Although we have focused on a specific architecture with Kerr nonlinearity, it should be anticipated that other architectures, which may include other types of nonlinearities (e.g. second order nonlinearities, or nonlinear absorbers), will give rise to a rich variety of mean-field behaviors in the presence of non-reciprocity.  

To explore the dynamics of noise, we consider the transient changes in the intensity of the lattice sites produced by perturbing the field in a single site from its steady-state value (Fig.~\ref{fig:nonreciprocal_noise_concept}). 
For a system with reciprocal couplings, such a perturbation would create disturbances which propagate in both directions across the lattice. 
However, the non-reciprocal couplings $(\kappa/g = 0.4)$ allow for a strikingly different behavior, in which fluctuations propagate in one direction, while decaying much more rapidly in the other (Fig. \ref{fig:nonreciprocal_noise_concept}d). 
This unidirectional propagation of fluctuations from steady-state arises directly from the non-reciprocity of \( H_N \), which is inherited from the non-reciprocal couplings in \( H^{(\text{lin})} \) of the parent lattice (see Eq.~\ref{eq:U}). 
We find that this effect persists across a broad range of system parameters and steady states, due to its connections to non-Hermitian topology, as we explore later.
This insensitivity to system parameters highlights the fundamental role played by non-reciprocity in shaping the noise dynamics. In what follows, we will explore the implications of this classical transient noise flow on the quantum noise properties of the system.

\begin{figure}
    \centering
    \includegraphics[width=\linewidth]{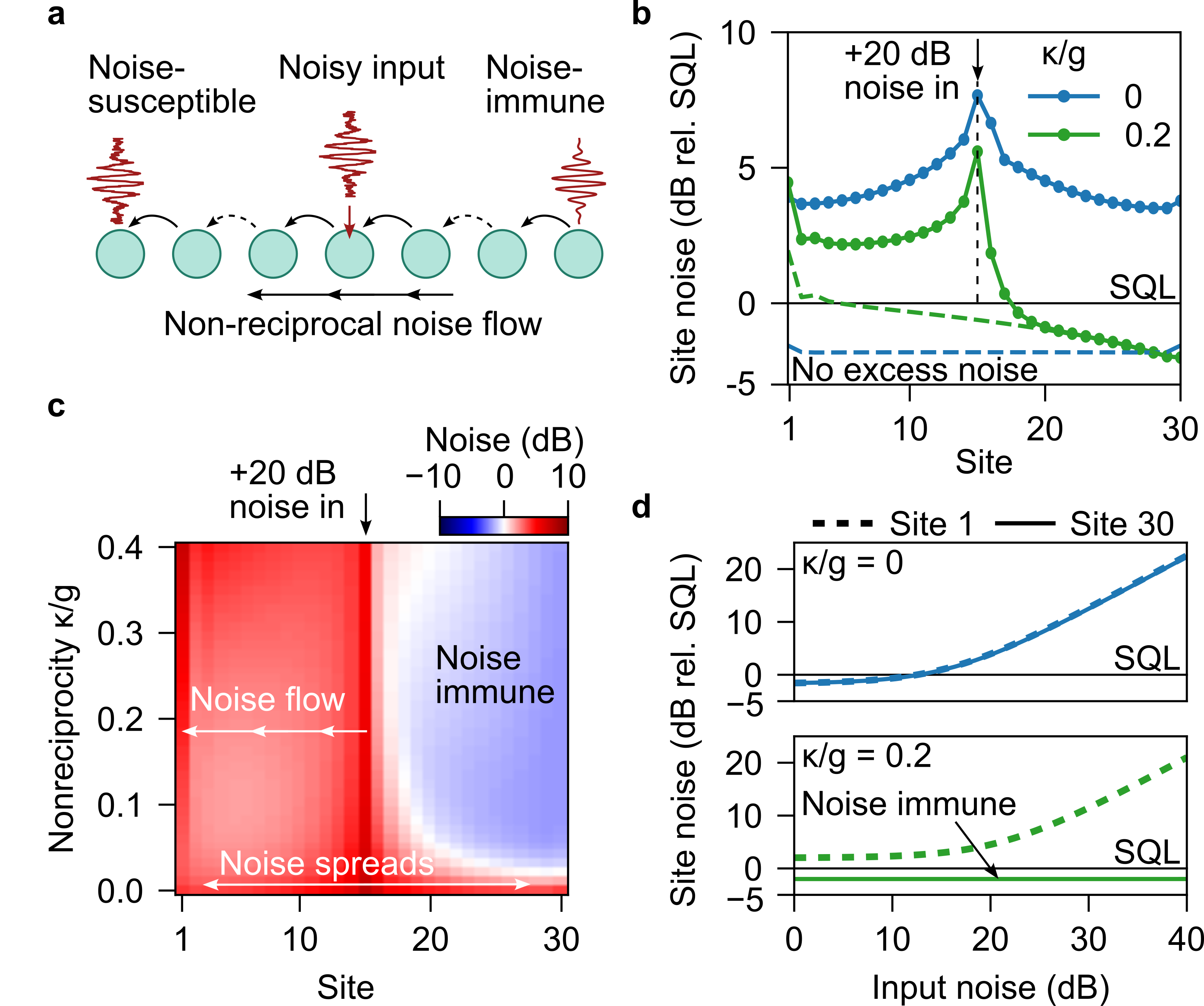}
    \caption{\textbf{Non-reciprocity enables immunity to external noise.} (\textbf{a}) Architecture for realizing noise-immunity through a non-reciprocal nonlinear lattice model based on the same type of cavity chain shown in Fig.~\ref{fig:nonreciprocal_noise_concept}. Excess noise input to some site in the middle of the system is routed in one direction due to non-reciprocity, leaving sites to the right strongly protected. (\textbf{b}) Steady-state intensity noise distribution for an identically driven open chain of Kerr resonators, where 20 dB of excess noise (phase-insensitive) is injected into the middle port, for both reciprocal $\kappa=0$ and non-reciprocal $\kappa/g=0.2$ systems. The dashed lines show the intensity noise levels without the excess noise input (i.e. for shot-noise-limited driving). We use nonlinear strength $\beta_i|s_i|^2/\omega_0^2 = 5$, limiting loss rate $\eta_i/\omega_0 = 2\times 10^{-2}$, linear coupling $g/\omega_0 = 0.2$, and external drives $s_i = 1$. (\textbf{c}) Steady-state intensity noise levels at each site as a function of non-reciprocity, in the presence of the same 20 dB excess noise shown in (b). Dashed lines show the values of $\kappa$ which are shown in (b). As the non-reciprocity increases, so does the degree of protection offered by the non-reciprocity, resulting in a noise-immune region of sites where even squeezing is retained in the presence of excess noise injection. (\textbf{d}) Noise levels of the boundary sites as a function of input noise level for a reciprocal ($\kappa=0$) and non-reciprocal ($\kappa/g = 0.2$) system. For the reciprocal system, the small degree of Kerr squeezing is quickly lost due to excess noise inputs, while the non-reciprocal system is immune to this deterioration.}
    \label{fig:noise_immunity}
\end{figure}

\section{Noise immunity through non-reciprocity} 
Now, we show that non-reciprocal noise transport results in the remarkable property of protection from excess noise injected into the system (Fig.~\ref{fig:noise_immunity}). 
Such excess noise is, for example, common in the high-powered amplified laser systems frequently employed to drive nonlinearities \cite{heffner1962fundamental}. 
Generically, excess noise injected into even a single channel of some system couples into all other channels, polluting them. 
On the contrary, we find that non-reciprocity can insulate one side of the system from added noise, providing total immunity to high noise levels (Fig.~\ref{fig:noise_immunity}a). 

To illustrate this, we consider a lattice of $N=30$ driven nonlinear resonators and compute the steady-state intensity noise distribution both with and without 20 dB of excess noise injected at the midpoint ($i=15$). 
In absence of excess noise, the Kerr nonlinearity at each site causes field quadratures to mix, leading to the generation of approximately 3 dB of intensity squeezing, as is known to arise in such driven nonlinear resonances \cite{drummond1980quantum}. 
In a reciprocal system, the excess noise propagates symmetrically, raising the noise level at all sites by at least 5 dB, and entirely degrading the existing squeezing (Fig.~\ref{fig:noise_immunity}b, blue curves).

However, in our non-reciprocal model, noise is funneled unidirectionally toward the left, preventing it from affecting the right side (Fig.~\ref{fig:noise_immunity}b, green curves).
While sites left of the noise source experience increased noise, those to the right remain unaffected. 
For the chosen parameters, resonators just a few sites away from the noise source exhibit no measurable impact. 
Remarkably, the rightmost sites maintain squeezing despite the presence of substantial excess noise, challenging the prevailing notion that noise necessarily degrades squeezing. 
Furthermore, the steady-state noise baseline (Fig.~\ref{fig:noise_immunity}b, green dashed line) exhibits an asymmetry due to non-reciprocal noise flow, a consequence of the non-Hermitian skin effect discussed in the next section.

This noise immunity emerges rapidly as non-reciprocity is introduced (Fig.~\ref{fig:noise_immunity}c). 
Even a modest non-reciprocity relative to the reciprocal coupling significantly protects the rightmost sites and restores squeezing. 
Increasing non-reciprocity further enhances this protection, nearly isolating all sites to the right of the noise source. 
Additionally, the sites at the right boundary remain immune even as the injected noise level increases (Fig.~\ref{fig:noise_immunity}d). 
For $\kappa/g=0.2$, the noise response of the rightmost site remains nearly constant across a 40 dB range of injected noise, maintaining a small degree of squeezing. 
In the S.I., we show that this effect is very resilient under disorder due to the fact that $H_N$ remains non-reciprocal for a wide range of system parameters and their corresponding steady-state field $\boldsymbol{\alpha}$. 

This effect is broadly compelling for multimode nonlinear systems, since this type of noise immunity can ensure robustness even as the number of nonlinearly interacting modes increases. 
Unlike sensitive mean-field dynamics, the noise behavior here becomes more predictable and rigid with scaling. 
This robustness is especially important for preserving the correlations present in multimode quantum states, which may play crucial roles in future architectures for metrology, data transmission, and quantum information processing.

\section{Non-Hermitian topology for noise} 
We next show that the unidirectional flow of noise can be understood from the the point-gap topology of the Hamiltonian governing noise dynamics, $H_N$.
We do so based on the topological classification schemes developed for non-Hermitian systems with complex eigenvalue spectra.~\cite{zhang2020correspondence, gong2018topological, okuma2020topological, bergholtz2021exceptional, wang2024non}. 
Under this scheme, an eigenvalue band is said to exhibit a point gap at a reference point in the complex plane if this point does not lie on the band. 
For one-dimensional periodic systems, the bands of eigenvalues of the linear equations generically form closed loops that enclose a finite area in the complex plane. 
Consequently, a band can exhibit a non-trivial winding around some fixed reference point, $\lambda_0 \in \mathbb{C}$, as captured by a winding number, $w(\lambda_0) \in \mathbb{Z}$:
\begin{equation}
    w(\lambda_0) = \int_{-\pi}^{\pi} \frac{\d k}{2\pi}\partial_k \arg(\lambda(k)-\lambda_0).
\end{equation}
Here, $\lambda(k)$ is the eigenvalue of the non-Hermitian Hamiltonian, $k$ is the wavevector associated with the one-dimensional periodicity with lattice constant set to unity. 

\begin{figure}
    \centering
    \includegraphics{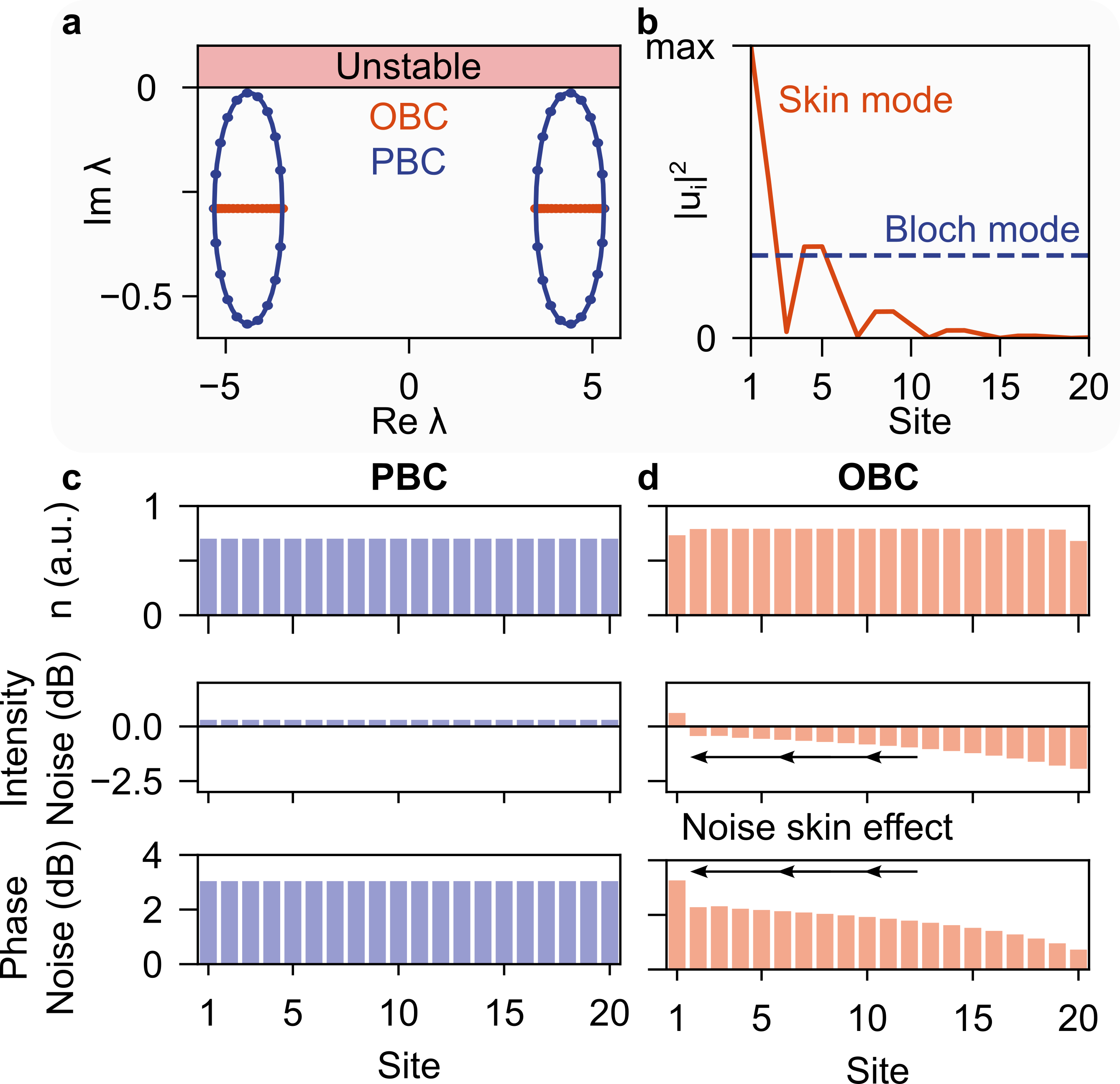}
    \caption{\textbf{Point gap topology and non-Hermitian skin effect for quantum noise.} Parameters used are $N=20$ resonators, linear coupling $g/\omega_0=0.4$, non-Hermiticity $\kappa/g = 0.3$, nonlinearity $\beta_i|s_i|^2/\omega_0^2 = 80$, and limiting loss $\eta/\omega_0 = 5\times 10^{-2}$. (\textbf{a}) Complex eigenvalues of the non-Hermitian noise Hamiltonian $H_N$ with periodic and open boundary conditions. (\textbf{b}) Typical eigenvectors under OBC/PBC, showing the contrast between Bloch modes (PBC), and edge-localized skin modes (OBC). The eigenvectors of the noise Hamiltonian $H_N$ take the form $(\mathbf{u}, \mathbf{u}^*)$, so that in the Figure, $u_i$ refers to the component at site $i$. (\textbf{c}) Steady state photon number ($n$) distribution, intensity noise, and phase noise, relative to shot noise, for the non-reciprocal system with PBC. (\textbf{d}) The same plots, but for the system with OBC. The system exhibits a non-Hermitian skin effect for quantum noise, resulting in asymmetric distributions of steady-state noise for both intensity and phase under OBC. This sensitivity to boundary conditions is reflected in the point-gap topology of $H_N$ shown in (a, b).}
    \label{fig:steady_state_noise}
\end{figure}

The topological nature of non-Hermitian systems often manifests through an extreme sensitivity to boundary conditions. 
Indeed, a major consequence of a non-trivial point-gap topology is that all eigenstates of a finite system, under open boundary conditions (OBC), are found to be localized near one of the boundaries, resulting in the so-called ``non-Hermitian skin effect'' (NHSE).
In the driven-dissipative context, the NHSE has previously been leveraged for directional amplification of fields~\cite{wanjura2020topological, wanjura2021correspondence, mcdonald2018phase, mcdonald2020exponentially}. 
In contrast, our focus is on a regime where strong nonlinearity suppresses chiral flows in the steady-state fields, yet the quantum noise exhibits a pronounced NHSE.

To show this, we consider nonlinear driven-dissipative Hatano-Nelson models with both periodic and open boundary conditions (PBC/OBC). 
We assume the detunings, drives, and losses to be identical for all sites. Under PBC, the steady-state field satisfies the equation $[\gamma_{\text{tot}}^2 + (\Delta + 2\beta n + 2g)^2]n = 2\eta|s|^2$, where $n = |\alpha|^2$ is the photon number which is identical for all sites, and $\gamma_{\text{tot}}\equiv \gamma + \eta$ is the total diagonal loss. 
Such a steady-state equation is known to enable bistability, where a single pump flux $|s|^2$ can support two steady intracavity intensities $n$ \cite{gibbs1981observation, gibbs1988optical, drummond1980quantum, lugiato1984ii}. 
The steady-states under different boundary conditions are very similar: uniform under PBC, and nearly uniform under OBC (Fig.~\ref{fig:steady_state_noise}c,d). 
However, the non-reciprocity strongly affects the properties of quantum noise, as we explain next.

For PBC, it is insightful to consider the fluctuations from the steady state in momentum space, defined as $\delta a_k = 1/\sqrt{N}\sum_n \delta a_n e^{-\im kn}$. 
In this basis, the fluctuations are easily diagonalized, and obey the Heisenberg-Langevin equation:
\begin{equation}
    \frac{\d}{\d t}\begin{pmatrix}
        \delta a_k \\ \delta a^\dagger_{-k}
    \end{pmatrix} = -\im H_N(k) \begin{pmatrix}
        \delta a_k \\ \delta a^\dagger_{-k}
    \end{pmatrix} + \begin{pmatrix}
        F_k(t) \\ F_{-k}^\dagger(t)
    \end{pmatrix}, 
    \label{eq:eom_k}
\end{equation}
where the noise Hamiltonian expressed in $k$-space is:
\begin{equation}
    H_N(k) = -\im\gamma_{\text{tot}} I + \im(\beta n)\sigma_y + (\Delta + 2\beta n + \omega_0(k)) \sigma_z.  
\end{equation}
Here, $\sigma_{y,z}$ are the Pauli matrices, $I$ is the identity matrix and $\omega_0(k) = 2g\cos k -2\im\kappa\sin k$. 
The Langevin correlations are given by $\braket{F_k(t) F_{k'}^\dagger(t')} = 2(\gamma_{\text{tot}} + 2\kappa\sin k)\delta_{kk'}\delta(t-t')$, which has the clear interpretation of a $k$-dependent loss which is different for $+k$ and $-k$ due to the non-reciprocity \cite{mcdonald2022nonequilibrium}.

The complex spectrum of $H_N(k)$ under PBC exhibits two loops with a non-trivial winding number of $w = +1$ for any reference point lying within the loops (Fig.~\ref{fig:steady_state_noise}a).
Under OBC, all eigenvalues of the finite system lie within these loops and the corresponding "noise" eigenvectors are localized to the left boundary of the system (Fig.~\ref{fig:steady_state_noise}b). 
Therefore, a noise source at any given site will couple into these eigenvectors, causing a preferential flow of noise towards the left boundary, which is the mechanism underlying the effect in Fig.~\ref{fig:noise_immunity}.

This unidirectional flow of noise results in a staircase-like distribution of both intensity and phase noise (Fig.~\ref{fig:steady_state_noise}d) under OBC, which can be understood in terms of the formation of skin modes of $H_N$ and the NHSE for quantum noise. 
Sites near the right boundary experience intensity squeezing from the Kerr nonlinearity, while sites near the left boundary accumulate excess noise in both intensity and phase due to the preferential leftward flow of fluctuations. 
Notably, these features are absent under PBC (Fig.~\ref{fig:steady_state_noise}c), where amplitude and phase noise are identical at each site due to the identical drives. 
This NHSE for noise generally survives under disorder, whether it enters through the system parameters or non-identical input drives, as long as the point gap remains open~\footnote{Note that since $H_N$ depends on the steady state, the connection between the PBC and OBC spectra in the context of the NHSE holds only when the steady states under both boundary conditions are nearly identical. Strong disorder can disrupt this by inducing different behaviors under PBC and OBC, such as significant differences in photon numbers, multistability, or instability in one of the cases. When this occurs, the NHSE of noise may be destabilized before the point gap closes.}.

Moving beyond the point gap topology, a richer topological classification is possible for non-Hermitian systems with multiple bands. 
In the present case, the Hamiltonian governing noise dynamics, $H_N$, contains doubled degrees of freedom corresponding to $\delta a_k$ and $\delta a^{\dagger}_{-k}$. 
As a result, in the simplest case of a single resonator per unit cell, $H_N$ hosts two bands under PBC (Fig.~\ref{fig:steady_state_noise}a). 
It is now known that the most general topological classification for a non-Hermitian system with $N$ bands is given in terms of the Braid group, $\mathbb{B}_N$~\cite{hu2021knots, wang2021topological, wojcik2020homotopy}. 
Such a classification characterizes the topologically distinct ways in which the bands wind around each other, forming braids in the space of the complex eigenvalues and the wavevector, $[ \mathrm{Re}(\lambda), \mathrm{Im}(\lambda), k ]$. 

\begin{figure}
    \centering
    \includegraphics{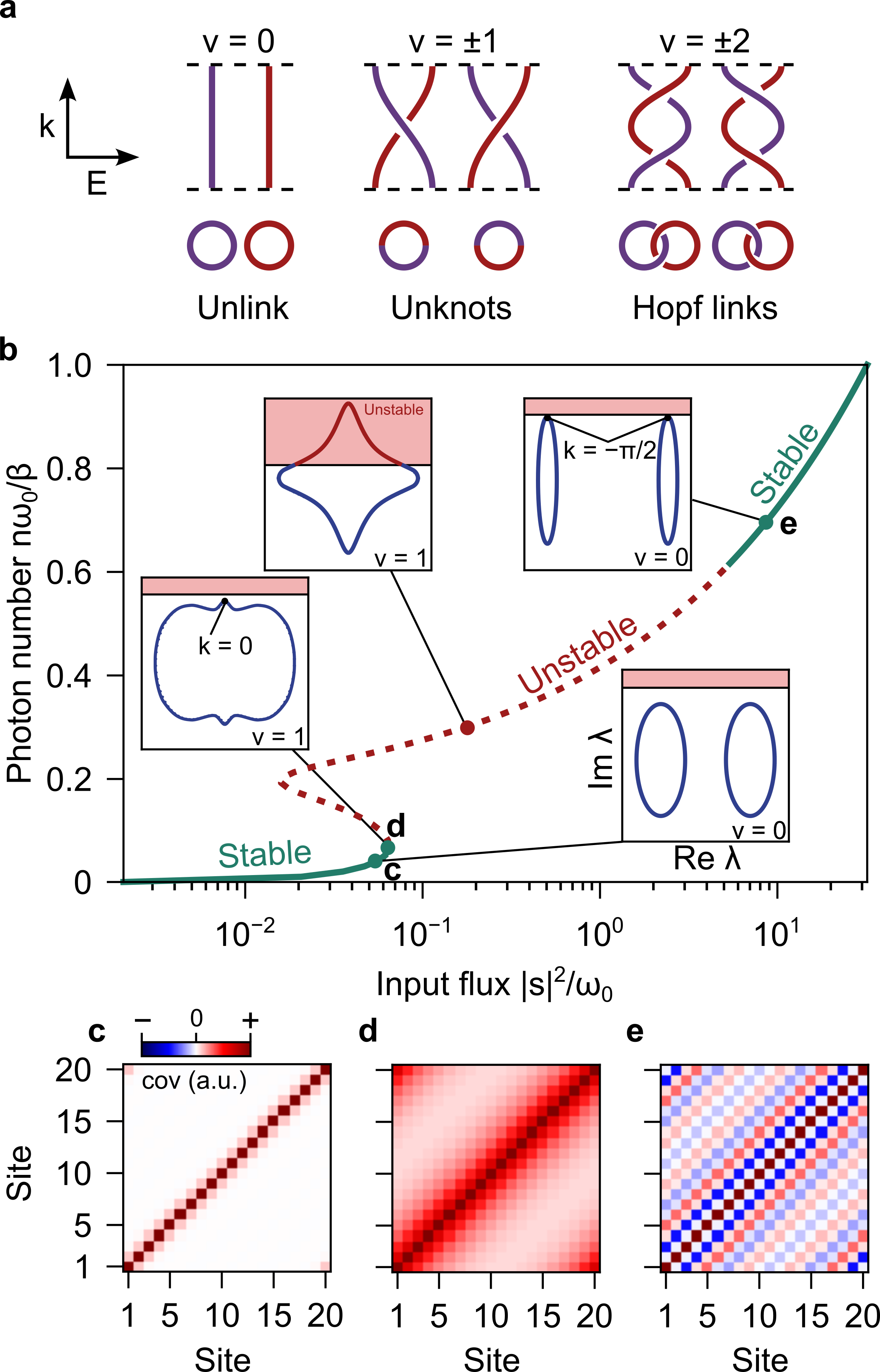}
    \caption{\textbf{Nonlinear phase diagram and correlations of non-reciprocal lattice model.} (\textbf{a}) Braid invariants $\nu$ associated with the braid group $\mathbb{B}_2$. (\textbf{b}) Phase diagram for the steady state photon number in each resonator as a function of input flux $|s|^2$ for a non-reciprocal nonlinear system under periodic boundary conditions. Stable and unstable regions are marked, and insets show the non-Hermitian winding curves based on the eigenvalues of $H_N(k)$ for $k$ in the first Brillouin zone. Parameters used are: linear coupling $g/\omega_0=0.05$, $\kappa/g = 0.3$, detuning $\Delta = -0.3$, limiting loss $\eta/\omega_0 = 10^{-2}$, nonlinearity $\beta|s|^2/\omega_0^2 = 1$.(\textbf{c-e}) Covariance of intensity fluctuations $\braket{\delta n_i \delta n_j}$ between sites $i$ and $j$ shown for the three marked points in panel (b) which have stable steady-state solutions. High levels of positively correlated noise are seen near the onset of instability (c). At higher photon numbers, solutions are stable again, and exhibit a pattern of positive and negative correlations due to the non-reciprocity (e).}
    \label{fig:phase_diagram}
\end{figure}

For the simplest non-trivial case of two bands, the group $\mathbb{B}_2$ provides a $\mathbb{Z}$ classification given by the braid degree, $\nu \in \mathbb{Z}$. 
For example, the $\nu = 0$ phase (unlink) is formed by two bands that do not braid around each other, while the $\nu = \pm 1$ phase (unknot) is formed by braiding once, and so on (Fig.~\ref{fig:phase_diagram}a). 
The transitions between the braid degrees are mediated by exceptional points in the spectrum~\cite{shen2018topological, ding2022non, hu2022knot}, resulting in some eigenvalues under OBCs forming complex conjugate pairs (up to a constant imaginary shift). 
Consequently, the OBC spectrum for $|\nu| > 0$ corresponds to noise eigenmodes with many different decay rates, which is distinct from the $\nu = 0$ case where the decay rates for all eigenmodes are identical due to the symmetries of the system. 
Therefore, the characterization of bands in terms of the braid degree complements the point-gap topology of individual bands and indicates whether or not this class of systems has a single time-scale for reaching equilibrium when noise is added.

Interestingly, we also find that these systems exhibit novel topological transitions in their noise spectra, mediated by the mean-field interactions between the bosonic noise degrees of freedom. 
In particular, we find that the braid degree of the bands of $H_N$ under PBC may change with the input flux of the drives due to the dependence of $H_N$ on the steady state fields. 
To demonstrate this with an example, we plot the steady-state photon number as a function of input flux of the drives and examine the braid degree of the bands for various points on this curve. 
We find that changing the input flux results in topological transitions from $\nu = 0$ to $\nu = 1$ and back to $\nu = 0$, mediated by exceptional points (Fig.~\ref{fig:phase_diagram}b). 
These exceptional point transitions can be related to the conditions for stability of the system to perturbations. 
For example, in one region of the phase space for our example system (e.g. between (c) and (d) marked in Fig.~\ref{fig:phase_diagram}b), the transition from $\nu = 0$ to $\nu=1$ pushes the eigenvalue at $k=0$ toward the real axis, eventually causing an unstable region where no steady-state exists.
We thus anticipate that this topological framework can aid the understanding of stability and noise in more general classes of nonlinear non-Hermitian lattices.

\section{Quantum correlations} 
In addition to inducing topological transitions in our system, nonlinearity can also facilitate the formation of quantum correlations between sites. 
One way to assess quantum correlations at various points of the phase diagram is by the covariance associated with intensity fluctuations between different sites in the periodic system (Figs.~\ref{fig:phase_diagram}b-d). 
In particular, we have used our theory to show that under PBCs, the covariance between some site $m$ and a site a distance $\ell$ away is given by:
\begin{equation}
\begin{split}
    \braket{\delta n_m \delta n_{m+\ell}} &= 
    \frac{|\alpha_m|^2}{N}\sum_k e^{\im k\ell} (\gamma_{\text{tot}} + 2\kappa \sin k) \\
    &\quad \times \int \frac{d\omega}{\pi} |\mu_k(\omega) + \nu_k(\omega)|^2.
\end{split}
\label{eq:correlations_k_space}
\end{equation}
Here, the frequency integral is taken over the $k$-dependent Bogoliubov coefficients defined so that $\delta a_k(\omega) = \mu_k(\omega)F_k(\omega) + \nu_{-k}^*(-\omega) F_{-k}^*(-\omega)$, which can be found by solving Eq.~\ref{eq:eom_k}. 
Positive covariance values indicate that intensity fluctuations from equilibrium tend to occur with the same sign, while negative values indicate fluctuations with opposing signs. 
The expression we have derived shows how these real-space correlations consist of a discrete Fourier transform of contributions from the $k$-resolved intensity noise levels. 
Thus, the spatial pattern of quantum correlation is largely determined by which wavevectors have the strongest fluctuations. 
Importantly, this relationship between real-space correlations and $k$-space noise allows us to connect the winding of the non-Hermitian noise bands (Fig.~\ref{fig:phase_diagram}a) to observed patterns in the spatial covariance (Figs.~\ref{fig:phase_diagram}c-e), which vary as a function of the input power level. 

At the first power level shown (Fig.~\ref{fig:phase_diagram}c), neighboring sites show some degree of positive covariance, owed to the nonlinear interactions between sites. 
Slightly higher power levels cause the system to approach an instability point. Here, the winding curve nearly touches the real axis at $k=0$. 
This directly results in high noise levels at $k=0$, which manifest as long-range positive correlation between sites (Fig.~\ref{fig:phase_diagram}d). 
As discussed above, the noise Hamiltonian $H_N$ passes through an exceptional point in between (c) and (d), causing the winding number $\nu$ to change just before the system becomes unstable. 

Finally, at a higher power (Fig.~\ref{fig:phase_diagram}e), the system supports a nontrivial and ordered state which is owed specifically to the non-reciprocity of the system. 
In the site basis, we observe a long-ranged ``checkerboard'' pattern of positive and negative covariance. In reciprocal space, this pattern results from strong noise at $k=\pm \pi/2$. 
These two $k$-points are exactly the points at which non-reciprocity causes the greatest disparity in the level of loss. 
Namely, $k=\pi/2$ has the greatest level of loss, while $k=-\pi/2$ has the lowest loss. 
For the parameters shown, the eigenvalues of $H_N(k=-\pi/2)$ nearly touch the real axis at the top of the winding ellipses (see inset of Fig.~\ref{fig:phase_diagram}a focused on point (e)), showing clearly how the winding diagram can be used to understand the correlations. 
This result is particularly notable as it demonstrates how, in a nonlinear system, the quantum noise sources intrinsic to non-Hermitian non-reciprocity can imprint themselves onto the quantum correlations of the driven system. 
In the future, this approach could open new avenues for generating topologically stabilized interacting states of light.

\section{Experimental discussion}
There are a variety of platforms which should be considered strong candidates for observing the effects in this work (and more broadly, studying the interplay between nonlinearity, non-reciprocity, and fluctuations from equilibrium). 
Here, we discuss a few of them, and also outline some of the approaches that could be used to measure these effects. 
In the S.I., we provide additional quantitative discussion of the relevant parameters for two example platforms.

One particularly effective way to create scalable lattices of photonic degrees of freedom is in the frequency domain. 
In this case, the field modes $a_i$ referred to in this work correspond to different optical resonances labeled by $i$. 
For example, there has been significant recent work on studying coupled frequency degrees of freedom in ``synthetic dimension'' lattices in fiber cavities \cite{wang2021generating, wang2021topological, senanian2023programmable}. 
In these systems, the lattice modes are the frequency modes in a fiber cavity, spaced out by the free spectral range (FSR) which is determined by the cavity length. 
Phase and amplitude modulations of the cavity at a frequency matching the FSR allows for both Hermitian and non-Hermitian couplings to be programmably induced between the frequency modes. 
In order to realize the quantum noise effects predicted in our work, one would need to operate such a system at intensities which are sufficiently high to see effects from Kerr nonlinearity in the fiber. 
Notably, such regimes of nonlinearity have been recently achieved in such fiber cavities which are driven by a few hundred mW of CW power \cite{leefmans2024topological, englebert2021parametrically, englebert2021temporal, englebert2023bloch, marques2023observation}.

In addition to fiber cavities, it could also be possible to implement a similar experiment using waveguide arrays \cite{sun2024photonic, schulz2021topological} or integrated platforms in materials such as thin film lithium niobate \cite{zhu2021integrated}, silicon carbide \cite{lukin20204h}, or silicon nitride \cite{moss2013new}, which achieve much stronger nonlinearities due to their modal confinements and high quality factors. 
In such resonators, nonlinearities can be triggered by injected powers which are mW scale or even lower. 
Progress in GHz rate modulators in this integrated setting suggest the possibility of exploring these programmable non-Hermitian coupling schemes in a controllable nonlinear setting (see, e.g. \cite{orsel2024giant}), providing a potential route to devices which exploit non-Hermitian topological schemes to robustly shape quantum states of multimodal light.

The most natural observable in these experiments is the noise level of different output frequency channels after the system has reached steady-state. 
The amplitude noise, for example, can be measured with a relatively simple setup involving a photodiode and an electrical spectrum analyzer which monitors the fluctuations of the photodiode current. 
With this type of measurement, non-reciprocal noise behavior can be assessed by measuring the noise for different frequency lattice sites, and confirming the type of asymmetry caused by the NHSE for noise. 
To test systems designed to provide the ``noise immunity'' through non-reciprocity that we have introduced, one can add gain (e.g. through EDFAs) to intentionally inject noisy light into an input channel, and then measure its effect on various output channels. 
We anticipate that fiber platforms are best suited for the first demonstrations of noise non-reciprocity in the nonlinear regime, due to their ease of programmability. 
In such systems, it should be possible to measure many dB of noise asymmetry due to the non-reciprocal noise flow across a nonlinear frequency lattice with modest injected powers.
Integrated nonlinear platforms, on the other hand, may be best suited for studies which aim to measure squeezing, due to the low losses which can be achieved on-chip. 

\section{Conclusion and outlook}
In summary, we have demonstrated that robust unidirectional flow of noise in multimode nonlinear quantum systems can be achieved through non-reciprocity introduced via non-Hermitian couplings between modes. 
We have also shown that this unidirectional flow leads to the immunity of certain parts of the system to excess input noise. 
These properties are a consequence of the non-Hermitian skin effect (NHSE), which arises from the point-gap topology of the Hamiltonian governing noise evolution. 
Additionally, we have demonstrated that these systems exhibit novel topological transitions mediated by nonlinearity, and that quantum correlations inherent to non-Hermitian systems appear here.

In the context of multimode nonlinear optics devices, our non-reciprocal approach to noise immunity can enable the removal of noise from amplified sources with minimal power loss, and create new architectures for nonlinear photonic quantum computation that are resistant to noise buildup in selected degrees of freedom. 
Furthermore, we have shown that non-reciprocity can be harnessed as a resource to generate quantum correlations, opening the door to new interacting phases of light supported by non-Hermitian systems. 
We have shown that a topological description is not only useful in predicting the properties of quantum noise but also reveals fundamentally new phenomena pertaining to the dynamics of noise in quantum systems. 
An interesting direction for future work is the characterization of these systems using real-space methods, which have been shown to be well suited for systems lacking periodicity~\cite{cerjan2022operator}, exhibiting strong nonlinearities~\cite{wong2023probing}, and non-Hermiticity~\cite{cerjan2023spectral, koekenbier2024transfer, garcia2024clifford}. 
Finally, we note that while we have focused on optical frequencies, the core concepts we have explored should be applicable in the context of other interacting wave systems, leading to potentially new directions in in microwave frequency devices (e.g. in electronics or superconducting citcuits \cite{nagulu2020non}), optomechanics \cite{peterson2017demonstration}, and acoustics \cite{nassar2020nonreciprocity}. 

\section{Acknowledgements}
The authors acknowledge fruitful discussions with Li Ge, Wladimir A. Benalcazar, Pedro Fittipaldi de Castro and Shiekh Zia Uddin. 
S.V. and M.S. acknowledge support from the U.S. Office of Naval Research (ONR) Multidisciplinary University Research Initiative (MURI) under Grant No. N00014-20-1-2325 on Robust Photonic Materials with Higher-Order Topological Protection. 
This work is also supported in part by DARPA Agreement No. HO0011249049 and the U. S. Army Research Office through the Institute for Soldier Nanotechnologies at MIT, under Collaborative Agreement Number W911NF-23-2-0121.
J.S. acknowledges previous support of a Mathworks Fellowship, as well as previous support from a National Defense Science and Engineering Graduate (NDSEG) Fellowship (F-1730184536). 
N.R. acknowledges the support of a Junior Fellowship from the Harvard Society of Fellows.

\bibliographystyle{unsrt}
\bibliography{references}
\end{document}